\documentclass[aip,apl,reprint]{revtex4-1}
\usepackage{graphicx}
\usepackage{dcolumn}
\usepackage{bm}

\usepackage{verbatim}

\usepackage{color}
\newcounter{mnotecount}[section]
\renewcommand{\themnotecount}{\thesection.\arabic{mnotecount}}
\newcommand{\mnotex}[1]
{\protect{\stepcounter{mnotecount}}$^{\mbox{\footnotesize
$
\bullet$\themnotecount}}$ \marginpar{
\raggedright\tiny\em 
$\!\!\!\!\!\!\,\bullet$\themnotecount: #1} }

\newcommand{\hl}[1]{{\color{black} #1}}

\begin{document}

\title{\hl{Measurement of single nanoparticle anisotropy by laser induced optical alignment and Rayleigh scattering for determining particle morphology}} 



\author{Markus Rademacher}
\author{Jonathan Gosling}
\author{Antonio Pontin}
\affiliation{Department of Physics \& Astronomy, University College London, London WC1E 6BT, United Kingdom}
\author{Marko Toroš}
\affiliation{School of Physics and Astronomy, University of Glasgow, Glasgow, G12 8QQ, United Kingdom}
\author{Jence T. Mulder}
\author{Arjan J. Houtepen}
\affiliation{Optoelectronic Materials Section, Faculty of Applied Sciences, Delft University of Technology, 2629 HZ Delft, The Netherlands}
\author{P. F. Barker}
\email[]{p.barker@ucl.ac.uk}
\affiliation{Department of Physics \& Astronomy, University College London, London WC1E 6BT, United Kingdom}



\begin{abstract}
We demonstrate the measurement of nanoparticle anisotropy by angularly resolved Rayleigh scattering of single optical levitated particles that are oriented in space via the trapping light in vacuum. This technique is applied to a range of particle geometries, from perfect spherical nanodroplets to octahedral nanocrystals. We show that this method can resolve shape differences down to a few nanometers and be applied in both low-damping environments, as demonstrated here, and in traditional overdamped fluids used in optical tweezers.  
\end{abstract}

\pacs{}

\maketitle 
The measurement of nanoparticle morphology is vitally important for aerosol science~\cite{zeller_direct_1985,xiong_morphological_2001,karlsson_numerical_2022}, nanoparticle production~\cite{sau_nonspherical_2010} and even identification of airborne viruses~\cite{gelderblom_structure_1996,ignatovich_optical_2006,pan_collection_2019,lukose_optical_2021}. For example, rapid measurement of the morphology of nanoparticles is of significance to determine their antibacterial activity and toxicity~\cite{sirelkhatim_review_2015}. Their geometry also determines their optical properties, which is important for cancer diagnosis and imaging~\cite{huang_gold_2010}. This also governs their specific absorption rate for in vivo applications of magnetic nanoparticle hyperthermia for cancer treatment~\cite{thanh_clinical_2018,sharma_nanoparticles-based_2019,hilger_vivo_2013}. Although accurate measurements can be carried out using electron microscopy or x-ray diffraction, light scattering methods can rapidly characterize single particles and even large ensembles in solution. However, as nanoparticles are often subject to translational and rotational Brownian motion, only the averaged properties of the particles can be measured. While this is sometimes sufficient, scattering from an aligned particle contains much more information when strongly suppressing Brownian motion.  

More recently, the field of levitated optomechanics has greatly progressed. It offers enhanced control over the translational and rotational motion of single isolated nanoparticles, and the movement can be strongly damped, suppressing Brownian motion. Particles are held in optical, electric, or magnetic traps. In optical traps, the particles can be localised in position down to a few picometers~\cite{pontin_simultaneous_2022,delic_cooling_2020,magrini_real-time_2021,tebbenjohanns_quantum_2021}. This control has allowed these systems to be brought into the quantum regime with cooling to the quantum ground state recently demonstrated~\cite{delic_cooling_2020,magrini_real-time_2021,tebbenjohanns_quantum_2021}. Such experiments have paved the way for the next generation of quantum applications ranging from quantum-limited sensing of gravity~\cite{rademacher_quantum_2020} to tests of the large-scale limit of quantum mechanics~\cite{bassi_models_2013}.
Knowledge of the shape, refractive index, and other properties of the levitated nanoparticles is critical for current and future quantum experiments~\cite{gonzalez-ballestero_levitodynamics_2021,stickler_quantum_2021}. These include quantum metrology which aims to search for new physics beyond the standard model~\cite{moore_searching_2021} and tests of quantum mechanics in this new mesoscopic regime~\cite{toros_creating_2021}.

A long-standing problem in levitated optomechanics is obtaining detailed information on the structure and geometry of the levitated nanometre-sized objects~\cite{van_der_laan_erratum_2021}. A basic procedure for determining the nanoparticle's shape has been developed based on linewidth measurements of the directional damping values for different spherical nanoparticle cluster configurations~\cite{ahn_optically_2018}.
While this technique is helpful, it is not very sensitive to small changes in particle shape within the nanometer range and cannot be used in the over-damped regime where many optical traps operate. 

Despite the importance of the laser scattering behavior of optically trapped nanoparticles to levitated optomechanics~\cite{tebbenjohanns_optimal_2019,li_fast_2021}, the measurement of the scattering pattern of an optically trapped object has not been used to determine the shape and geometry of a single particle. In this study, we determine single nanoparticle morphology using laser-induced optical alignment of levitated nanoparticles coupled with angularly resolved Rayleigh scattering. We present the underlying laser Rayleigh scattering theory on which the characterization technique is based and demonstrate its application to a range of trapped and oriented symmetric top nanoparticles with different morphology. These results are compared with numerical simulations of the particle alignment. We demonstrate that this method effectively determines the geometry of the optically levitated particle, realizing a new tool for studying isolated nanoparticles.

\begin{figure}[h]
\centering
\includegraphics[width=\columnwidth]{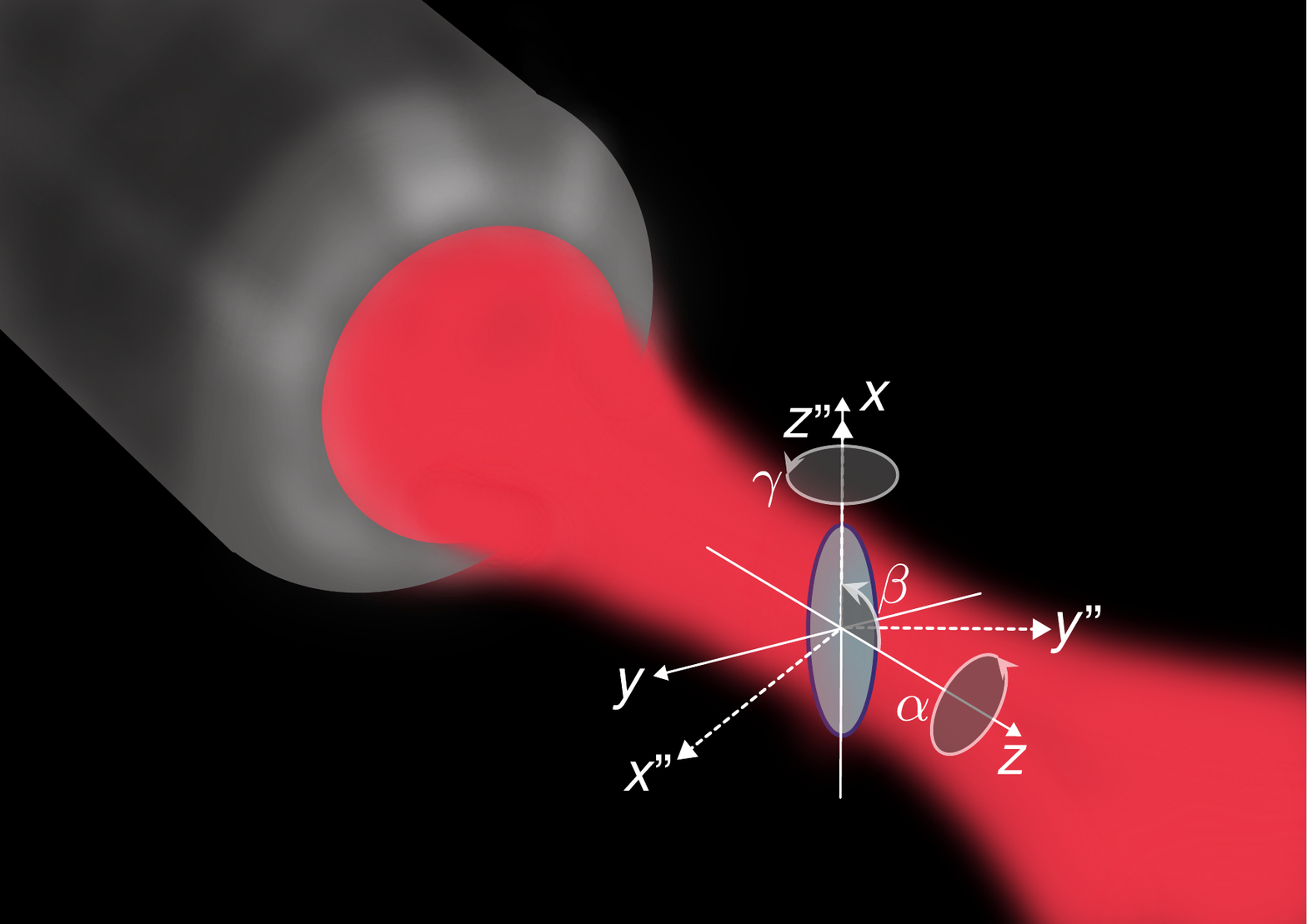}
\caption{Schematic representation of an ellipsoidally shaped nanoparticle trapped in an optical tweezer. The three Euler angles ($\alpha$, $\beta$ and $\gamma$) and three centers of mass coordinates are shown in the convention used throughout the work (laboratory frame: $x$, $y$ and $z$; body frame: $x"$, $y"$ and $z"$).}
\label{fig:schematic_rot}
\end{figure}

A key ingredient in this method is the ability to align the nanoparticle to the laboratory reference frame via the optical torque induced by the linearly polarized trapping light. This feature greatly simplifies the treatment of the light scattering and increases its sensitivity to particle shape because it allows us to minimize the orientational averaging.

We consider light scattered from a linearly polarised laser along the y-axis as shown in Figure~\ref{fig:schematic_rot}, and we measure the vertically polarised light intensity $I_V$ scattered by our nanoparticle along the x-axis. We treat the sub-wavelength nanoparticle as a dipole scatterer~\cite{miles_laser_2001}. The incoming electric field is horizontally polarised with $E_{\text{Iy}}=\sqrt{2 \frac{I_y}{c \epsilon _0}}$ and the scattered intensity of the light along the x-axis is given by:
\begin{equation}
    \label{equ:IV}
    I_V=\frac{\pi ^2 c E_{\text{Iy}}^2 \chi_{\text{yy}}^2}{2 \lambda ^4 r^2 \epsilon _0},
\end{equation}
where $c$ is the speed of light, and $\epsilon_0$ is the vacuum permittivity. 

To determine the asymmetry and the geometry of an optically trapped nanostructure from its Rayleigh scattering~\cite{miles_laser_2001}, we determine the susceptibility $\chi_{yy}$ for any angle of the particle with respect to the observation direction as determined by the polarisation and intensity of the levitation field. The electric susceptibility tensor $\bm{\chi}$ of the nanoparticle in the body frame, when aligned with the laboratory frame, is given by
\begin{equation}
\label{equ:chi0}
  \bm{\chi_0} = 
\left(
\begin{array}{ccc}
 \chi_{\text{0}}^{zz} & 0 & 0 \\
 0 & \chi _{\text{0}}^{yy} & 0 \\
 0 & 0 & \chi _{\text{0}}^{xx} \\
\end{array}
\right)
\end{equation}
To predict the scattered light in our observation direction for any orientation of the nanoparticle, we calculate the susceptibility tensor by 
\begin{equation}
    \label{equ:BodyToLab}
    \bm{\chi}=\bm{R}\bm{\chi_0}\bm{R}^\mathsf{T}.
\end{equation}
The angles shown in Figure~\ref{fig:schematic_rot} are the Euler angles which represent the three rotations of the principal nanoparticle axes $\alpha$, $\beta$ and $\gamma$, where $\bm{R}$ is the rotation matrix using the $z$-$y'$-$z"$ convention~\cite{arfken_mathematical_1993} and is given by $\bm{R}_z(\alpha) \bm{R}_y(\beta) \bm{R}_z(\gamma)$ as shown in Appendix B.
The rotation matrix $\bm{R}$ transforms the susceptibility matrix from the body frame aligned with the lab frame to the orientation determined by the trapping laser. {Combining equations~\ref{equ:IV}-\ref{equ:BodyToLab} we can extract $\bm{\chi_0}$ by placing the nanoparticle in different orientations.}

To illustrate how the scattered light changes for a non-spherical nanoparticle as a function of its orientation, we consider a levitated prolate spheroid for which an analytical solution exists for the susceptibility. The particle is an ellipsoid, as shown in Figure~\ref{fig:schematic_rot}, where two out of the three radii ($r_x, r_y, r_z$) are of the same length, and the diagonal components of the susceptibility tensor are given~\cite{bohren_absorption_1998}:
\begin{equation}
    \label{equ:chi_ellipsoid}
   \chi_{0}^{\{zz,yy,xx\}} = \frac{4 \pi  r_x r_y r_z}{\epsilon_0 V} \frac{\left(\epsilon _l-\epsilon
   _m\right)}{3 L_{\{1,2,3\}} \left(\epsilon _l-\epsilon
   _m\right)+3 \epsilon _m} 
\end{equation}
where $L_1=\left(1-e^2\right) \left(\frac{\log \left(\frac{e+1}{1-e}\right)}{2 e}-1\right)/e^2 $, $e=\sqrt{1-\frac{r_y^2}{r_x^2}}$ and that $L_2=L_3$
with $L_1+L_2+L_3=1$. 

\begin{figure*}[!ht]
    \centering
    \includegraphics[width=6.69in]{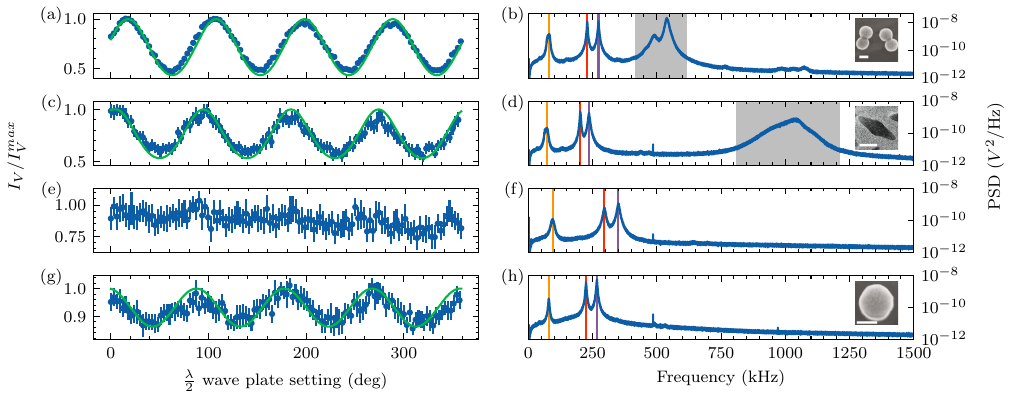}
    \caption{Experimental scattering and time trace data: (a), (c), (e), (g) show normalized scattering intensity data $I_V/I_V^{max}$ versus $\lambda/2$ wave plate setting, i.e., half the particle orientation angle $\alpha$ in free space. (b), (d), (f), and (h) show power spectral densities of the time traces of the nanoparticle's motion recorded in the PBS-detection setup. \hl{The orange, red and purple lines represent the oscillation frequency of the z, x and y degree of freedom. The grey shaded area in (b) and (d) denote the $\alpha$ and $\beta$ degrees of freedom, which are non-linearly coupled by the $\gamma$ motion~\cite{seberson_parametric_2019}.} All datasets in (b), (d), (f) and (h) corresponds to the same samples used for the data in (a), (c), (e) and (g). (a) and (b) is data for a nanometre-sized dumbbell structure of two nanospheres attached. The scale bar in insets in (b), (d) and (h) represents 100~nm. The inset in (b) shows a scanning electron micrograph (SEM) of the nanodumbbells. The dataset in (c) and (d) corresponds to a nanometre-sized bi-pyramidal structure with a short length of (77$\pm$5)~nm and a long side of (227$\pm$18)~nm. {The inset in (d) displays a SEM of the octahedron structure.} (e) and (f) displays data collected for a liquid methanol droplet suspended in the optical trap. (g) and (h) represent the data collected for a nanometre-sized sphere from the same sample (r=(71$\pm$2)~nm from microparticles GmbH). The inset in (h) exhibits a SEM of the nanoparticle.}
    \label{fig:scattering_results}
\end{figure*}

Using equations~\ref{equ:IV}-\ref{equ:chi_ellipsoid}, we calculate the intensity modulation as the angle $\alpha$ is varied over $2 \pi$ with $\beta=\pi/2$ and $\gamma=0$ fixed for different nano-ellipsoids with different degrees of asymmetry as defined by the ratio of the lengths $r_x/r_y$ and where $r_y=r_z$. {$\gamma$ can be set to any value for axially symmetric particles without changing the results.}
\begin{figure}
    \centering
    \includegraphics[width=\columnwidth]{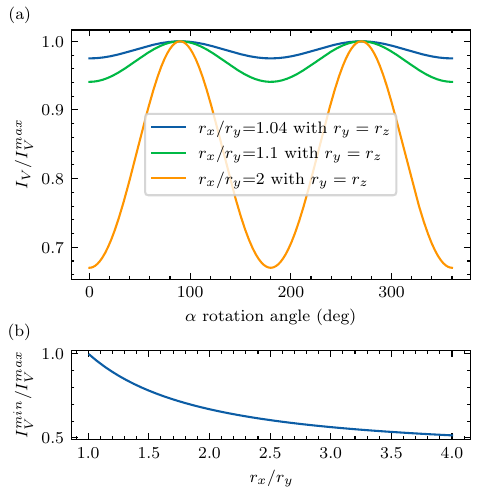}
    \caption{(a) Relative scattering intensity of the vertical polarised light scattered by the ellipsoid $I_V/\text{max}\left(I_V\right)$ versus the rotation angle $\alpha$ depicted in Figure~\ref{fig:schematic_rot}. The blue line represents an ellipsoid with 4\% asymmetry. The green one shows a 10\% difference for the directional radii, and the orange has 100\% asymmetry. (b) Total modulation depth of the scattering intensity of the vertical polarised light scattered by an ellipsoid $\text{min}\left(I_V\right)/\text{max}\left(I_V\right)$ versus the asymmetry parameter of the different radii of a prolate spheroid $r_x/r_y$ where $r_y=r_z$.}
    \label{fig:IV_ellipsoid}
\end{figure}
 These calculations are shown for three ellipsoidal nanoparticles with different asymmetry ratios in Figure~\ref{fig:IV_ellipsoid}(a). We define the total intensity modulation depth as $\text{min}\left(I_V\right)/\text{max}\left(I_V\right)={I_V\left(\bm{\chi_0}\right)}/{I_V\left(\bm{\chi_{\alpha=\frac{\pi}{2}}}\right)}$, which is plotted as a function of the asymmetry parameter $r_x/r_y$ for different ellipsoids (prolate spheroids) in Figure~\ref{fig:IV_ellipsoid}(b). {We consider the angular Brownian motion of the particle when it is optically aligned. This alignment depends on the optical field, the susceptibility, and the gas temperature. This process acts to reduce the observed asymmetry in the angular scattering. Computationally we first calculate the full scattering pattern of the particle and perform averaging on this pattern using the calculated angular dynamics of the optically trapped particle assuming a harmonic trapped motion and applying the equipartition theorem~\cite{arita_coherent_2020,pontin_simultaneous_2022}. We apply this averaging process to the scattering patterns shown in Figure~\ref{fig:IV_ellipsoid} and the solid green lines in Figure~\ref{fig:scattering_results}. For all our calculations, we use the laboratory temperature of $295$~K and vacuum pressure of 5~mbar.} The figures show that the larger the asymmetry parameter, the larger the intensity modulation depth of the vertical polarised scattered light. The modulation visible in figures~\ref{fig:IV_ellipsoid}(a) and~\ref{fig:IV_ellipsoid}(b) are size independent and only depend on the asymmetry ratio of the nanoparticle. This dependence on particle asymmetry is non-linear and has enhanced sensitivity when small changes in the asymmetry ratio exist. 
 Similar behavior is seen for other morphologies. When an analytical description of the susceptibility tensor~$\bm{\chi}_0$ is not available, we calculate it numerically as described in Appendix A.  

\begin{figure}
    \centering
    \includegraphics[width=0.8\columnwidth]{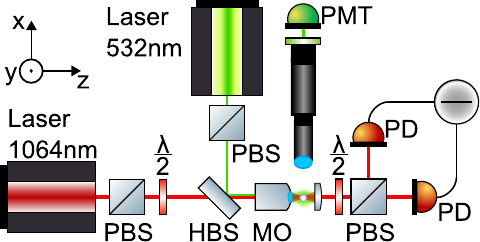}
    \caption{Schematic of the experimental setup for characterizing the shape and geometry of levitated nanoparticles by analyzing their scattered light modulation. $\lambda_{\text{trap}}=1064$~nm laser light in red, $\lambda_{\text{probe}}=532$~nm probing laser light, in green.}
    \label{fig:exp_setup}
\end{figure}
We demonstrate this angularly resolved Rayleigh scattering on different nanoparticles and verify its sensitivity to asymmetry. Figure~\ref{fig:exp_setup} shows the experimental set-up that was used. Here, the trapping light ($\lambda_{\text{trap}}=1064$~nm\hl{, P$_{\text{trap}}=450$~mW}) was polarised by a polarising beam splitter (PBS) and passed through a $\frac{\lambda}{2}$-waveplate, which allowed us to rotate the polarization. Linear polarization rotation was achieved by mounting it in a remote piezo-controlled kinematic rotary mount. A green probe laser ($\lambda_{\text{probe}}=532$~nm\hl{, P$_{\text{probe}}=5$~mW}) of fixed linear polarisation was Rayleigh scattered from the aligned particle. This laser illuminated the trapped nanoparticle by coupling it into the trapping beam path via a harmonic beam splitter (HBS).
Both co-linear beams were passed into a microscope objective ($\text{NA}_{\text{trap}}=0.8$) and focused. The strongly focused 1064~nm beam trapped the particle, and a lens collected the transmitted light ($\text{NA}_{\text{coll}}=0.77$). The light was subsequently passed through a PBS where the two arms are aligned onto two separate photodiodes implementing a balanced detection scheme. We used a $\frac{\lambda}{2}$-wave plate to balance the power on the two photodiodes. This detection scheme allows us to record a time trace of all six degrees of freedom of the levitated nanoparticle since the interaction of the motion of the nanoparticle with the trapping light changes the polarisation state and direction of the laser light~\cite{ahn_optically_2018,reimann_ghz_2018,monteiro_optical_2018,tebbensjohanns_optimal_2022,toros_detection_2018}. The green probe beam was linearly polarized along the y-direction of Figure~\ref{fig:schematic_rot}. This beam was apertured when coupled into the microscope objective to create a much larger focal spot at the trapped nanoparticle. The larger focus allows us to reduce the light intensity on the particle so that there are no significant optical forces on it, and the light provides uniform illumination of the particle. The $532$~nm scattered light from the nanoparticle was collected along the x-axis by a lens system with an effective focal length of $3.5$~cm. The collected light passed through a $532$~nm narrow bandpass filter and was detected on a photomultiplier tube (PMT). 

The nanoparticles studied \hl{were loaded into the chamber by creating an aerosol using an asthma nebulizer~\cite{burnham_holographic_2006,millen_optomechanics_2020}.} They were trapped at atmospheric pressure and the pressure was then reduced to $5$~mbar. The vacuum environment allowed us to compare our results with the dynamical motion of the particle in the under-damped regime. A time trace of the balanced detector for rotation was recorded as shown in Figure~\ref{fig:exp_setup}. The power spectral density (PSD) of this time trace was calculated and used to confirm the optical trapping of non-spherical particles and the alignment via their librational motion. This PSD also shows the translational motion of the particle in the trap.
We record the scattered light from a fixed direction as a function of the particle's orientation in free space. At the same time, we change the particle's orientation by rotating the trapping beam linear polarization. To do this, we rotate the half-wave plate in 120 equal increments over 360°. This waveplate rotation turns the particle 720°. The PMT voltage allows to determine the angularly resolved Rayleigh scattering pattern.

The angularly resolved scattering and the PSDs for a range of levitated nanoparticles are shown in Figure~\ref{fig:scattering_results}. The data includes nanodumbbells of silica, octahedrons of yttrium lithium fluoride, spheroidal nanoparticles of silica, and spherical methanol nanodroplets. The normalised scattered light intensity data in Figure~\ref{fig:scattering_results}(a), (c), (e), (g) corresponds to the power spectral densities shown in Figure~\ref{fig:scattering_results}(b), (d), (f), (h).

The data set in Figure~\ref{fig:scattering_results}(a) and (b) represents a silica dumbbell consisting of two r=(71$\pm$2)~nm spheres stuck together. The data set in Figure~\ref{fig:scattering_results}(a) shows the distinct modulation of the scattering intensity upon turning the nanodumbbell in the 532~nm detection beam. This modulation agrees well with the calculation shown in the solid green line in Figure~\ref{fig:scattering_results}(a). We also consider the angular Brownian motion for this calculated scattering modulation, which reduces the overall contrast in the modulation depth. We can distinctly distinguish the librational motion of the nanodumbbell from the PSD of the rotational motion in Figure~\ref{fig:scattering_results}(b). 

Figure~\ref{fig:scattering_results}(c) is the scattering modulation recorded for an octahedron-shaped nanoparticle with a short side length of $a=(77\pm5)$~nm and a height of $h=(113.5\pm9)$~nm (Appendix A). This modulation also agrees with the simulations shown as a solid green line. The power spectral density for the motion of the octahedron-shaped structure is shown in Figure~\ref{fig:scattering_results}(d). The librational peak {at $\sim1$~MHz} is higher in frequency than the librational peaks of the nanodumbbell sample in Figure~\ref{fig:scattering_results}(b). This difference is consistent with the simulations of the optical trapping setup. 

The data set shown in Figure~\ref{fig:scattering_results}(e) and (f) represents an optically levitated pure methanol droplet. This droplet is a near-perfect sphere due to high tension forces. Within the uncertainty of our measurements, this is confirmed by the scattering data set in Figure~\ref{fig:scattering_results}(e), which shows no consistent modulation of the scattered intensity. In addition, the power spectral density shown in Figure~\ref{fig:scattering_results}(f) exhibits no librational peaks in their spectrum, which is also a sign of their sphericity.

Lastly, we present in Figure~\ref{fig:scattering_results}(g) the corresponding scattering data sets, which shows a smaller modulation of the scattering intensity{, compared to Figure~\ref{fig:scattering_results}(a) and~\ref{fig:scattering_results}(c),} for commercially available silica nanospheres (r=(71$\pm$2)~nm from microparticles GmbH) while being rotated in the $532$~nm illumination field. Nevertheless, we cannot identify any distinct peak of the librational motion in the PSD shown in Figure~\ref{fig:scattering_results}(h). However, we see an increased noise background in the frequency range between the lowest and second highest translational motion peak, which is not seen in any particle morphology pointing to a lower frequency in the optically induced alignment of the silica particles. The scattering modulation in Figure~\ref{fig:scattering_results}(g) indicates that the nanosphere deviates from sphericity by 15 percent with an asymmetry ratio of $r_x/r_y=0.85$. This divergence is equivalent to a change of 10~nm in radius between one axis and another. This difference is calculated by fitting the experimental scattering data with equation~\ref{equ:IV}. The procedure was repeated for several trapped nanospheres, and all show some asphericity derived from the modulation in their angular scattering pattern. For clarity, these are not shown in Figure~\ref{fig:scattering_results}(g) but range from 0.84 to 0.86 in the asymmetry ratio. From the signal-to-noise ratio demonstrated \hl{($\Delta I_V/I^{max}_V=0.061$) in Figure~\ref{fig:scattering_results}(g) in conjunction with the nonlinear function in Figure~\ref{fig:IV_ellipsoid}(b)}, we could differentiate changes between $r_x$ and $r_y$ of approximately 5~nm \hl{for a nanosphere of $r=71$~nm}. The ability to measure this small difference using the scattering method indicates its utility for fast measurements of particle morphology. Such a small asphericity cannot be determined from the PSD as no librational peak can be observed at our pressures.
Additionally, recent work~\cite{pontin_simultaneous_2022} in our laboratory suggests that most of the time, when loading spherical nanoparticles, we do not trap perfect spheres but rather trap nanoparticles that have the shape of an ellipsoid with a small degree of anisotropy.

In this paper, we showed that the modulation in the angularly resolved scattering of an optically suspended nanoparticle is a valuable tool to characterize small asymmetries in single nanoparticles down to the percent level. This method was demonstrated on a range of non-spherical nanoparticles and is possible due to the control in the alignment of the nanoparticle in the optical field and the strong asymmetry in the angular scattering when aligned. Although we have demonstrated this technique in an under-damped gas environment, it is also applicable for use with traditional over-damped tweezers experiments carried out in liquids for biological samples. 
In future experiments cooling the angular motion degree of freedom of the sample using feedback polarisation control of the trapping light or by feedback optical torque control~\cite{bang_five-dimensional_2020} or by cavity cooling~\cite{pontin_simultaneous_2022} would reduce the effects of angular Brownian motion, which would enhance the modulation contrast and increase our ability to detect even more minor nanoparticle asymmetries. Trapping and cooling based on elliptically polarised light rather than the linearly polarised light here would allow this technique to be applied to other morphologies, including asymmetric tops, i.e., arbitrary shape characterization of nanoparticles. 

In addition, further improvements in the detection signal-to-noise ratio would allow us to differentiate length differences as small as 0.4~nm. We expect this method could be used with other scattering and imaging modalities, such as super-resolution imaging, as part of a correlated imaging approach~\cite{fonta_correlative_2015} to nanoparticle imaging and analysis. \hl{Another exciting application for this technique is characterization of compound nanoparticles with a complex internal structure~\cite{chen_nanochemistry_2016}. For example, particles used for drug delivery have anisotropic scattering due to internal polycrystal structure~\cite{noskov_non-mie_2018}, shape-induced depolarization~\cite{bahrom_controllable_2019}, and filling with dopants~\cite{noskov_golden_2021}. Our technique holds promise for distinguishing these contributions which are important for using nanoparticles in biophotonic applications. Whilst
the modulation depth shown in Figure~\ref{fig:scattering_results} measures optical anisotropy of the
aligned nanoparticle, a measurement of the absolute amplitude provides an
additional step towards determining particle morphology.}
\hl{Furthermore, this technique is not limited to the Rayleigh regime as discussed here, but can be further extended to larger particles using multi-polar Mie scattering theory~\cite{bohren_absorption_1998}. Larger particles up to $32~\mu$m have been shown to levitate and align using counter-propagating and gravito-optical traps~\cite{monteiro_optical_2017}.} This work also has important implications for more fundamental studies in levitated optomechanics, where a detailed knowledge of particle morphology is required to compare experiments with theoretical predictions \cite{blakemore_precision_2019}.
\\


M.R., J.G., A.P. and P.F.B. acknowledge funding from the EPSRC Grant No. EP/S000267/1 and EP/W029626/1. M.R., J.G., A.P., J.T.M., A.J.H and P.F.B. acknowledge funding from the H2020-EU.1.2.1 TEQ project Grant agreement ID: 766900. M.T. acknowledges funding by the Leverhulme Trust (RPG-2020-197). 

\section*{References}
\bibliographystyle{apsrev4-1}
\bibliography{references.bib}


\end{document}


\title{Supplementary Material: Measurement of single nanoparticle anisotropy by laser induced optical alignment and Rayleigh scattering for determining particle morphology} 



\author{Markus Rademacher}
\author{Jonathan Gosling}
\author{Antonio Pontin}
\affiliation{Department of Physics \& Astronomy, University College London, London WC1E 6BT, United Kingdom}
\author{Marko Toroš}
\affiliation{School of Physics and Astronomy, University of Glasgow, Glasgow, G12 8QQ, United Kingdom}
\author{Jence T. Mulder}
\author{Arjan J. Houtepen}
\affiliation{Optoelectronic Materials Section, Faculty of Applied Sciences, Delft University of Technology, 2629 HZ Delft, The Netherlands}
\author{P. F. Barker}
\email[]{p.barker@ucl.ac.uk}
\affiliation{Department of Physics \& Astronomy, University College London, London WC1E 6BT, United Kingdom}

\maketitle
\setcounter{equation}{0}
\setcounter{figure}{0}
\setcounter{table}{0}
\setcounter{page}{1}
\makeatletter
\renewcommand{\figurename}{FIG.}
\renewcommand{\theequation}{S\arabic{equation}}
\renewcommand{\thefigure}{S\arabic{figure}}
\renewcommand{\bibnumfmt}[1]{[S#1]}
\renewcommand{\citenumfont}[1]{S#1}

\section*{Appendix A - Calculation of the susceptibility tensor for arbitrary shaped particles}
\begin{figure}[!ht]
\centering
\includegraphics[width=0.1\textwidth]{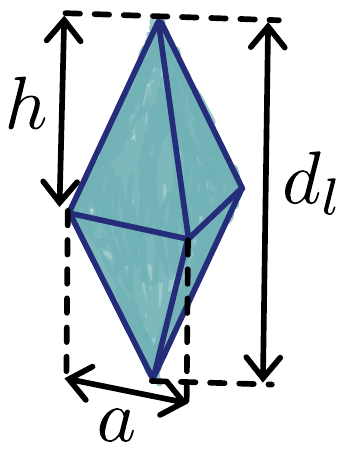}
\caption{Schematic illustration of YLF octahedron nanoparticle with length measurements indications}
\label{fig:octahedron}
\end{figure}
To compare our experiment with theory we numerically calculate the three directional susceptibilities $\chi_{zz}$, $\chi_{yy}$ and $\chi_{xx}$ of non-ellipsoidal objects where analytical descriptions are not available. To do this we numerically calculate the scattered intensity in the Rayleigh regime (where the particle dimensions is significantly less than the wavelength). We use the Rayleigh scattering relations~\cite{miles_laser_2001} to determine the susceptibility matrix $\bm{\chi_0}$. To calculate this we determine the intensity of the vertical polarized scattered light $I_V$ of the nanometre sized object for a constant horizontally polarized incoming electric field using the finite-difference time-domain (FDTD) method~\cite{yee_numerical_1966,taflove_application_1980}. We determine the scattered intensity for the trapping wavelength $\lambda_{\text{trap}}=1064$~nm at a distance of $r=1$~m and express the susceptibility $\chi_{yy}$ as in equation 1 of the main text. Equation 1 of the main text provides access to $\chi_{zz}$ by rotating the object in reference to the incoming light polarization along the z-axis~\cite{miles_laser_2001}. With $\chi_{zz}$ now obtained, we also calculated $\chi_{xx}$ due to the symmetric properties of the nanodumbbell and irregular octahedron nanoparticles. Our numerical calculations for nanodumbbells are in very good agreement with reference~\cite{ahn_optically_2018}. 

With the directional susceptibilities of an irregular octahedron and a nanodumbbell now calculated we determine the susceptibility matrix $\bm{\chi_0}$ and calculate the angular trapping dynamics for $\lambda_{\text{trap}}=1064$~nm. In Figure 4(a) and (c) of the main text the relative vertical polarized scattered light intensity modulation $I_V/I^{max}_V$ versus the $\frac{\lambda}{2}$wave plate angle is shown as a green solid line calculated for $\lambda_{\text{probe}}=532$~nm using the FDTD method without assuming any scattering regime. The modulation in graph 4(e) of the main text is calculated for an irregular octahedron with a short axis length of $a=(77\pm5)$~nm and a height of $h=(113.5\pm9)$~nm, i.e. a long axis length of $d_l=(227\pm18)$~nm and has an asymmetry ratio $d_l/a = 2.95$. Please see Figure~\ref{fig:octahedron} for a schematic illustration of the octahedron shape.

\section*{Appendix B - Rotation matrix}
$\bm{R}_i$:

\begin{equation}
\label{equ:R1}
  \bm{R}_z(\xi) =  \left(
\begin{array}{ccc}
 \cos (\xi ) & -\sin (\xi ) & 0 \\
 \sin (\xi ) & \cos (\xi) & 0 \\
 0 & 0 & 1 \\ 
\end{array}
\right),
\end{equation}

\begin{equation}
\label{equ:R2}
   \bm{R}_y(\xi) = \left(
\begin{array}{ccc}
 \cos (\xi ) & 0 & \sin (\xi ) \\
 0 & 1 & 0 \\
 -\sin (\xi ) & 0 & \cos (\xi ) \\
\end{array}
\right),
\end{equation}

\begin{equation}
\label{equ:R3}
\bm{R}_x(\xi) = \left(
\begin{array}{ccc}
 1 & 0 & 0 \\
 0 & \cos (\xi ) & -\sin (\xi ) \\
 0 & \sin (\xi ) & \cos (\xi )  \\
\end{array}
\right).
\end{equation}

{
\section*{Appendix C - colloidal synthesis of YLF nanocrystals}
YLF nanocrystals are produced by a colloidal synthesis method where complexes of Li-oleate, and Y-oleate react with fluoride by cracking of trifluoroacetate at temperatures above 250~$^{\circ}$C. This results in the formation of YLF nuclei, which grow into bipyramidal shaped nanocrystals. This synthesis yields monodisperse, bipyramidal, nanocrystals that are colloidally stable with a much higher reproducibility. The oleate surfactants can be removed and replaced by BF4-. This prevents unwanted light absorption by the organic ligands and renders the colloidal particles stable in methanol.}
\section*{References}
\bibliographystyle{apsrev4-1}
\bibliography{references.bib}